\documentclass{article}

\usepackage{PRIMEarxiv}
\usepackage{xcolor}   
 
\usepackage[utf8]{inputenc} % allow utf-8 input
\usepackage[T1]{fontenc}    % use 8-bit T1 fonts
\usepackage{hyperref}       % hyperlinks
\usepackage{url}            % simple URL typesetting
\usepackage{booktabs}       % professional-quality tables
\usepackage{amsfonts}       % blackboard math symbols
\usepackage{nicefrac}       % compact symbols for 1/2, etc.
\usepackage{microtype}      % microtypography
\usepackage{orcidlink}
\usepackage{lipsum}
\usepackage{fancyhdr}       % header
\usepackage{graphicx}       % graphics
\graphicspath{{media/}}     % organize your images and other figures under media/ folder
\usepackage{amsmath}
\title{Single-ended Recovery of Optical Fiber Transmission Matrices using Neural Networks}

\author{Yijie Zheng\orcidlink{0000-0002-6513-1584}\textsuperscript{1},Terry Wright\textsuperscript{1},Wen Zhong\textsuperscript{2,3},Qing Yang\textsuperscript{2,3},George S. D. Gordon\orcidlink{0000-0002-7333-5106}\textsuperscript{1*}\\
\textsuperscript{1}Optics and Photonics Research group, University of Nottingham, UK\\
\textsuperscript{2}State Key Laboratory of Extreme Photonics and Instrumentation, College of Optical Science and Engineering\\ International Research Center for Advanced Photonics, Zhejiang University, Hangzhou, China\\
\textsuperscript{3}Research Center for Humanoid Sensing, Zhejiang Lab, Hangzhou, China\\
\textsuperscript{*}Corresponding Author:  \texttt{george.gordon@nottingham.ac.uk}}

\begin{document}
\maketitle

\begin{abstract}
Ultra-thin multimode optical fiber imaging promises next-generation medical endoscopes reaching high image resolution for deep tissues. However, current technology suffers from severe optical distortion, as the fiber's calibration is sensitive to bending and temperature and thus requires in vivo re-measurement with access to a single end only. We present a neural network (NN)-based approach to reconstruct the fiber's transmission matrix (TM) based on multi-wavelength reflection-mode measurements. We train two different NN architectures via a custom loss function insensitive to global phase-degeneracy: a fully connected NN and convolutional U-Net. We reconstruct the 64 $\times$ 64 complex-valued fiber TMs through a simulated single-ended optical fiber with $\leq$ 4\% error and cross-validate on experimentally measured TMs, demonstrating both wide-field and confocal scanning image reconstruction with small error. Our TM recovery approach is 4500 times faster, is more robust to fiber perturbation during characterization, and operates with non-square TMs.

\end{abstract}
% keywords can be removed
\keywords{Optical fiber imaging  \and Transmission matrix reconstruction \and Custom loss function \and Neural network}

\section{INTRODUCTION}
Ultra-thin endoscopes are a promising technique for enabling cell-scale imaging in difficult-to-reach parts of the body, with the potential to improve disease detection in organs such as the pancreas and ovaries. Commercial products using imaging fiber bundles around 1mm in diameter are used in bile ducts \cite{BostonScientific} and flexible and full-color imaging has been demonstrated using distal scanning mechanisms that are typically around 2mm in diameter \cite{Lee2010,untracht2021imaging,hwang2020handheld}. To further reduce the size of endoscopes, recent work has focused on imaging through ultra-thin multimode fibers with diameters of 0.125mm and has achieved \emph{in vivo} fluorescence imaging in brains of immobilized mice \cite{turtaev2018high}. However, there are some key limitations of these imaging systems that use ultra-thin optical fiber. First, the thinnest such imaging devices are made using multimode fiber (MMF), which suffers from optical distortion that changes whenever the fiber is perturbed, particularly for longer fibers ($>$1m) required to reach deep inside the human body \cite{Psaltis2016}. Second, to calibrate this distortion, practical endoscopes made using MMF or some types of fiber bundle require measurement of their transmission matrix (TM) which requires transmittg a set of well-defined modes of light from the proximal facet to the distal facet where the resulting optical field must be measured. If calibration is required immediately before use (e.g for \emph{in vivo} use), such components would be required on the distal tip and would thus compromise the ultra-thin form factor \cite{Gu2015}.

\begin{figure}[!htpb]
    \centering
    \includegraphics{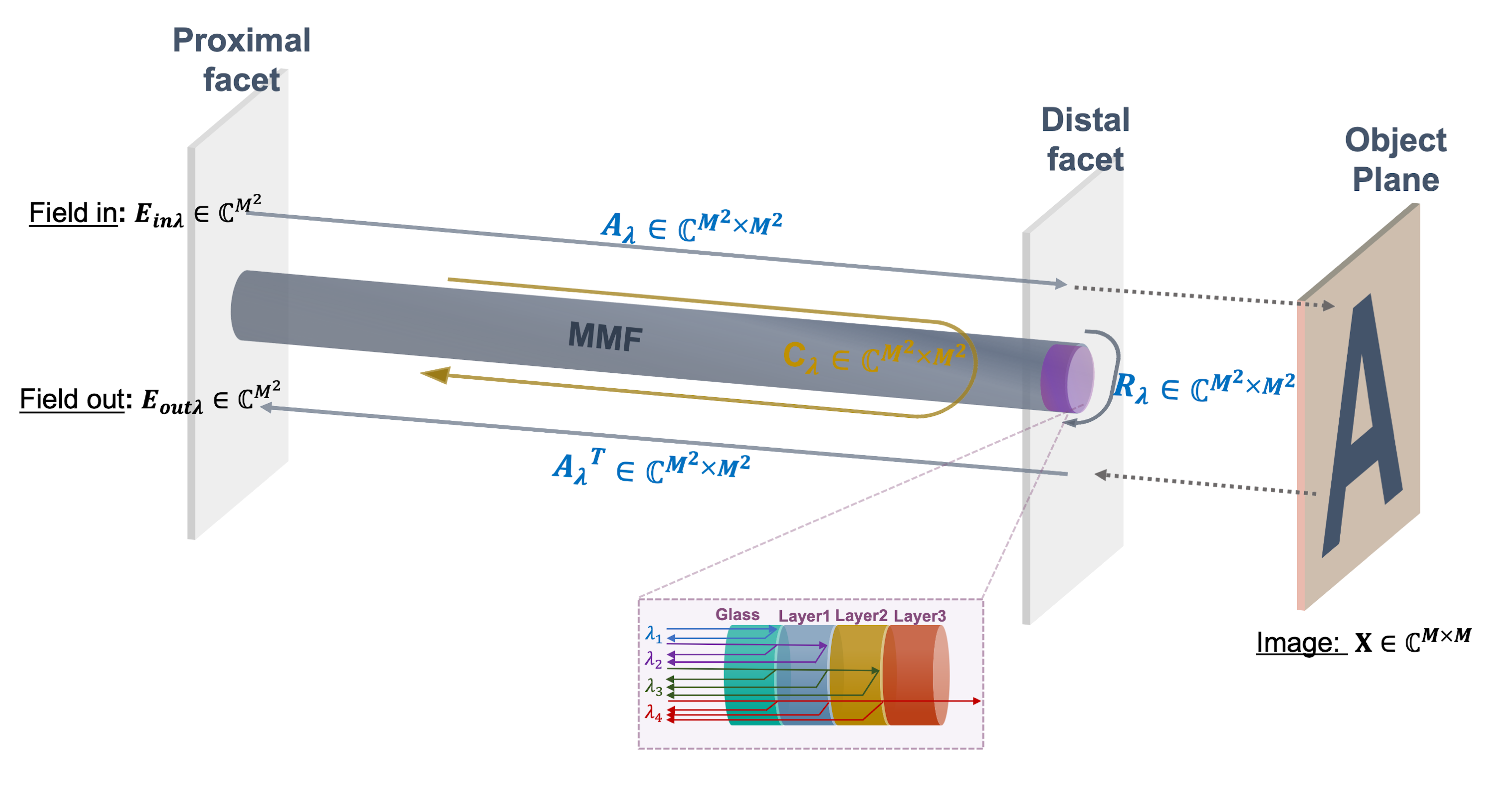}
    \caption{Single-ended optical fiber imaging system for transmission matrix (TM) recovery. The object, which will produce an image $\mathbf{X}\in\mathbb{C}^{M\times M}$ is placed at the distal facet. Light with a field of $\mathbf{E_{in}}\in\mathbb{C}^{M^2}$ propagates from the proximal facet through the optical fiber, with the forward TM of the optical fiber defined as $\mathbf{A_{\lambda}}\in\mathbb{C}^{M^2\times M^2}$ at the wavelength, $\lambda$. A reflector stack with a three-layer structure is placed at the distal facet, producing a reflector matrix $\mathbf{R_{\lambda}}\in\mathbb{C}^{M^2\times M^2}$ at the wavelength $\lambda$. There are four different wavelengths used, where wavelengths $\lambda_1$, $\lambda_2$, and $\lambda_3$ are used for characterization and $\lambda_4$ for imaging. For each wavelength, light propagates through one or more layers of the stack and is partially reflected by metasurfaces at the interfaces between layers. This produces distinct reflector matrices at each wavelength, $R_{\lambda_{1..3}}$, which can be addressed using a tuneable laser. At wavelength $\lambda_4$ light passes fully through the stack so that imaging can be performed. Faithful image reconstruction require correction for the TM of the reflector stack using pre-calibrated values at this wavelength \cite{gordon2019characterizing}. Reflection matrices, $\mathbf{C_{\lambda}}\in\mathbb{C}^{M^2\times M^2}$, can be measured at relevant wavelengths to recover the TM.} 
    \label{fig:intro}
\end{figure}

A number of methods have been proposed to calibrate fiber TMs without distal access including guidestars \cite{li2021memory,li2017focus,weiss2018two}, beacons that can be tracked \cite{farahi2013dynamic, wen2023single}, or reflective structures on the fiber tips \cite{Gu2015,chen2020remote,gordon2019characterizing}. Gordon et al. \cite{gordon2019characterizing} proposed a single-ended method of TM recovery based on the fiber system shown in Figure \ref{fig:intro}, with a specially designed reflector stack that provides different reflectances at different wavelengths. This approach avoids the need for measurement at both proximal and distal end of the fiber and works for non-unitary TMs. The reflection matrix, $\mathbf{C_{\lambda}}\in\mathbb{C}^{M^2\times M^2}$, where images are assumed to be $M\times M$ pixels, describes how an incident field $\mathbf{E_{in}}\in\mathbb{C}^{M^2}$ is transformed via propagation through the optical fiber, reflected by the reflector stack and finally transferred back through the fiber into an output field $\mathbf{E_{out}}\in\mathbb{C}^{M^2}$ at a wavelength of $\lambda$:
\begin{equation}
    \mathbf{C_{\lambda}} \mathbf{E_{in\lambda}} = \mathbf{E_{out\lambda}}
\end{equation}
The reflectance matrix, $\mathbf{C_{\lambda}}\in\mathbb{C}^{M^2\times M^2}$, is obtained from a forward pass via the TM, $\mathbf{A_{\lambda}}\in\mathbb{C}^{M^2\times M^2}$, reflection via the reflector stack, $\mathbf{R_{\lambda}}\in\mathbb{C}^{M^2\times M^2}$, and then a return pass via the TM, $\mathbf{A_{\lambda}}^{\top}\in\mathbb{C}^{M^2\times M^2}$. It has been previously been shown that the forward TM at wavelength $\lambda$, ,$\mathbf{A_{\lambda}}$, can be unambiguously reconstructed based on the measured reflection matrices at 3 different wavelengths, with imaging performed at a fourth wavelength \cite{gordon2019characterizing}.  This requires design of a special reflector stack whose reflection matrix changes with wavelength at a significantly faster rate than the fire TM, which may be achieved by alternately stacking metasurface reflectors with long-pass filters. A first-order dispersion model is assumed for the TM and has been experimentally validated to be $>70\%$ accurate over a 5nm bandwidth for a 1-2m length MMF using 110 modes, avoiding degeneracies arising from matrix logarithms \cite{gordon2019characterizing}, although recent work has shown MMF dispersion modeling over a much greater bandwidth \cite{lee2023efficient}. This leads to a set of 3 non-linear quadratic-form equations:
\begin{equation}    \mathbf{C_{\lambda_1}}=\mathbf{{A}_{\lambda_1}^T}\mathbf{R_{\lambda_1}}\mathbf{A_{\lambda_1}}
\label{eqn2}
\end{equation}
\begin{equation}    
\mathbf{C_{\lambda_2}}={(e^{(\log\mathbf{{A_{\lambda_1}}}\frac{\lambda_1}{\lambda_2})})}^T\mathbf{R_{\lambda_2}}\mathbf(e^{(\log\mathbf{{A_{\lambda_1}}}\frac{\lambda_1}{\lambda_2})})
\label{eqn3}
\end{equation}
\begin{equation}    
\mathbf{C_{\lambda_3}}={(e^{(\log\mathbf{{A_{\lambda_1}}}\frac{\lambda_1}{\lambda_3})})}^T\mathbf{R_{\lambda_3}}\mathbf(e^{(\log\mathbf{{A_{\lambda_1}}}\frac{\lambda_1}{\lambda_3})})
\label{eqn4}
\end{equation}
\noindent where $\mathbf {e^{(\log\mathbf{{A_{\lambda_1}}}\frac{\lambda_1}{\lambda_2})}}$ is the TM adjusted for a wavelength $\lambda_2$. 

Currently, these equations are solved by using an iterative approach which relies on optimization of the entire TM \cite{gordon2019characterizing}. This therefore scales in complexity with the square of the matrix dimension, incurring significant computational time, especially for large matrices. In practice, the TM shows high sensitivity to bending and temperature so in a practical usage scenario would need to be measured very frequently and reconstructed immediately prior to imaging. Large computational times are therefore not practical.

Considering this, there are several methods that have been developed in order to reduce the computational time for fiber imaging. These methods typically exploit prior knowledge about the fibers to improve or speed up TM reconstruction.  For example, Li et al. \cite{li2021compressively} proposed a compressed sampling method based on the optical TM to reconstruct full-size TM of a multimode fiber supporting 754 modes at compression ratios down to 5\% with good fidelity. Similarly, look-up tables can be used with a reflective beacon for high-speed TM estimation over an experimentally sampled prior space \cite{wen2023single}. Huang et al. \cite{Huang2020} retrieved the optical TM of a multimode fiber using the extended Kalman filter, enabling faster reconstruction. 

Recently, there has been work on using deep learning approaches, involving convolutional neural networks, to reconstruct images via multimode fibers both in transmission and reflection modes \cite{rahmani2018multimode,liu2023single,fan2022deep}. These methods have the advantage of being fast, and also learning and utilising prior information about the fiber properties and the objects being imaged.  However, their performance typically degrades significantly under fiber perturbation because they do not have access to reflection calibration measurements required to unambiguously resolve a TM.  Further, because such approaches seek to approximate the forward propagation of light and often only consider amplitude image recovery, they often rely on classical mean-squared error loss functions for training.

In order to incorporate reflection calibration measurements following fiber perturbation, it may instead be advantageous to use AI approaches to reconstruct a TM rather than an image, though there has been relatively little work in this area.  When reconstructing a TM comprising complex numbers, a particular type of degeneracy arises that is not well handled by conventional AI loss functions: a global phase factor.  In many physical problems, including the recovery of TMs for the purposes of image reconstruction and phase-hologram generation, global phase factors are not relevant as they do not affect the perceived performance of the system: it is the \emph{relative} phase between pixels that must be preserved. The relative phase between rows or columns of fiber TMs or reflection matrices is typically preserved using a reference beam e.g. interferometry \cite{mouthaan2022robust} or referencing to a fiber mode \cite{gordon2019full}. Global phase may have a physical interpretation related to the physical length of the fiber, but in practice it is often arbitrary unless great care is taken.  For example in interferometric systems the global phase is likely to be arbitrary unless the optical path lengths of the reference and sample arms are perfectly matched, which is very challenging for multimode fibers.  Further, the global phase often drifts during practical experiments \cite{mouthaan2022robust}, and approaches using phase-retrieval produce entirely arbitrary global phase values \cite{gordon2019full}. Therefore, in many practical situations, conventional loss functions will convert arbitrary shifts in the global phase to large changes in loss, which can confound minimization algorithms used to fit AI models and cause overfitting.

In this paper, we therefore present a method of implementing single-ended recovery of an optical fiber TM by solving Eqn. \ref{eqn2}-- Eqn. \ref{eqn4} based on three reflection matrix measurements at three different wavelengths. Specifically, we present two different neural network architectures, fully connected neural network (FCNN) and convolutional U-net based neural networks, and demonstrate the performance of both. As a necessary step, we account for the global phase factor of the entire TM by developing a custom global phase insensitive loss function that avoids degeneracies introduced by conventional loss functions such as mean-squared error (MSE). We first validate our model by recovering $64\times64$ complex-valued fiber TMs through a simulated single-ended optical fiber system (shown in Figure \ref{fig:intro}) with $\leq 4\%$ error for both FCNN and convolutional U-net architectures. We then demonstrate image reconstruction through fiber based on recovered TMs for two different imaging modalities: widefield imaging, achieving $\leq 9\%$ error, and confocal imaging, achieving $\leq 5\%$ error. We highlight several advantages of this TM recovery approach compared to previous TM recovery methods. Firstly, once the model is trained ($\sim$100 hours), it only requires $\sim$1 second for reconstruction, which is 4500 times faster than pre-existing iterative approaches. Secondly, the conventional method \cite{gordon2019characterizing} can only reconstruct square TM cases, whereas this method is compatible with non-square-shaped TM with $\leq 8\%$ error, useful for many practical cases where optical systems may have different mode bases at proximal and distal ends. Third, the requirements for the reflectors in terms of matrix properties are related and then can have arbitrary distributions of eigenvalues.

\section{RESULTS}
\subsection{Simulated TM recovery}
\begin{figure}[!htpb]
    \centering
    \includegraphics{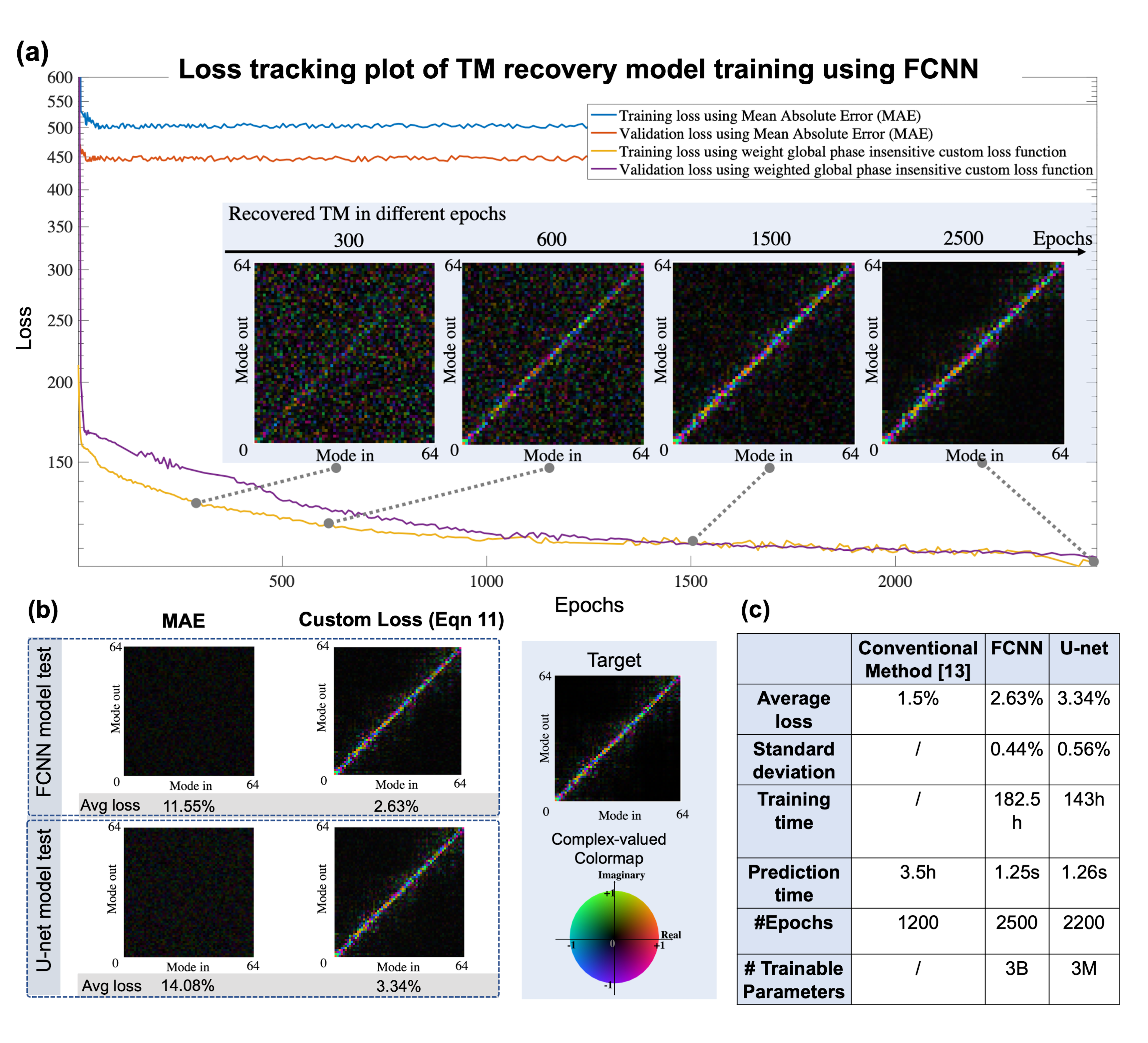}
    \caption{Simulated TMs recovered by training Neural Network (NN) models. (a) Training and validation loss tracking plot using MAE and custom loss function (Eqn. \ref{eqn:loss3}) of fully-connected neural network (FCNN) model. Blue and red line represent training and validation loss using mean absolute error (MAE) loss function. Yellow and purple line represent training and validation loss using our global-phase insensitve loss function. TM recovery results over different epochs are shown inset. (b) TM recovery results using two different neural network architectures (i.e. FCNN and convolutional U-net networks), with two different loss functions, namely MAE and custom loss function in Eqn. \ref{eqn:loss3}. (c) Comparison between FCNN and convolutional U-net architecture in aspects of average loss, standard deviation, training time, prediction time, number of converging epochs and number of trainable parameters. In complex-valued colormap, hue represents phase and lightness/darkness represents amplitude.}
    \label{fig:Simulated_TMrec}
\end{figure}

This TM recovery model was trained on a simulated dataset comprising 900,000 sets of simulated reflection matrices, $\mathbf{C}_\lambda$ at 3 wavelengths, $\lambda_1 = 850$nm, $\lambda_2 = 852$nm and $\lambda_3 = 854$nm, as input and a complex-valued non-unitary TM at wavelength $\lambda_1$, $\mathbf{A}_{\lambda_1}$ as output. Performance was monitored during training using 200,000 validation data that were not used in training. Final performance figures for the model were evaluated using an additional 100,000 data that were not part of the validation set. Figure \ref{fig:Simulated_TMrec}(a) shows the training and validation loss in training the FCNN model over 2500 epochs using different loss functions, namely conventional mean absolute error (MAE), and our global-phase insensitive custom loss function (Eqn.\ref{eqn:loss3}). Our global-phase insensitive loss functions show a decreasing loss in both training (yellow line) and validation (purple line) in the first 2000 epochs and convergence after 2500 epochs, whereas the MAE loss function exhibits fluctuating non-converging loss values for both training (blue line) and validation (red line). An example of a reconstructed TM predicted by the FCNN model at different epochs is shown inset in Figure \ref{fig:Simulated_TMrec}(a). It can be seen that the predicted TM approaches the target TM from 300 epochs to 2500 epochs, enabled by the custom loss function.

Figure \ref{fig:Simulated_TMrec} (b) compares the TM result predicted by our two different neural network architectures using different loss functions. Both FCNN and convolutional U-net-based neural networks cannot recover TM when using the MAE loss function but are capable of recovering TM using the global phase insensitive loss function, with an average loss (metric defined in Eqn. \ref{eq:averageloss}) of $\leq 4\%$ over 100,000 test TMs. Compared to the previous iterative approach, which requires nearly 10 hours to converge with an average loss of $\leq 1.5\%$, our method shows a larger $\leq 4\%$ average loss, but images retain acceptable visual quality of reconstructed images (shown in Result Section Widefield image reconstruction based on recovered TM). Furthermore, we also evaluated the computational resource usage of the two different neural network architectures as shown in Figure\ref{fig:Simulated_TMrec} (c). Training was implemented using Tensorflow 2.0 running on an NVIDIA Tesla V100 GPU. Compared to FCNN, the convolutional U-net shows significant advantages in memory usage, requiring 1000 times fewer trainable parameters, and in convergence time, which is reduced by $20\%$. However, it shows $0.7\%$ larger average loss on the test set. Both FCNN and convolutional U-net can recover TM at a loss $\leq 4\%$ (standard deviation 0.44\% and 0.56\% respectively), with $\sim$1s prediction time.
\subsection{Widefield image reconstruction based on recovered TM}
\label{subsec:widefieldResults}
\begin{figure}[!htpb]
    \centering
    \includegraphics{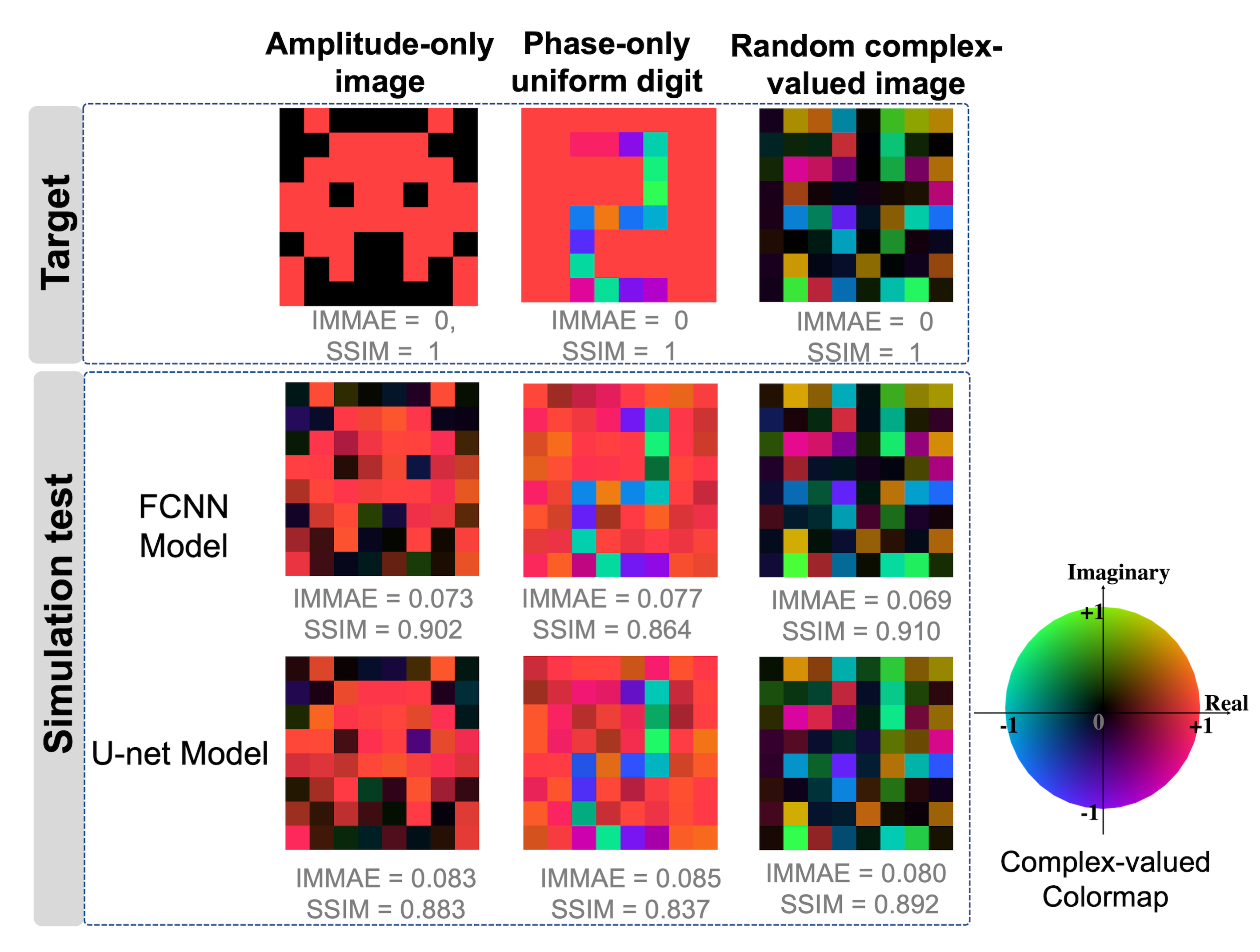}
    \caption{Widefield image reconstruction based on recovered transmission matrices (TMs) using fully-connected neural network (FCNN) and convolutional U-net networks. We considered 3 example
    images: an amplitude-only image with a ‘space invader’ pattern, a phase-only digit with a uniform
    amplitude and a random complex-valued image. Two metrics are defined to evaluate the image reconstruction performance, namely image mean absolute error, IMMAE (Eqn. \ref{eq:widefield_IMMAE}) and structural similarity index measure, SSIM (Eqn. \ref{eq:widefield_SSIM}).In complex-valued colormap, hue represents phase and lightness/darkness represents amplitude.}
    \label{fig:Widefield_imaging}
\end{figure}

Next, we examine a widefield imaging modality to evaluate the performance of recovered TMs at a fourth wavelength (the imaging wavelength), $\lambda_4=854.5$nm, where we assume a pixel basis at the fiber output. We considered 3 example images denoted $\mathbf{x}\in\mathbb{C}^{8\times 8}$: an amplitude-only image with a `space invader' pattern, a phase-only digit with a uniform amplitude and a random complex-valued image. Random noise was added to better simulate expected behaviour in a real system (shown in Eqn. \ref{eq:imaging}). Figure \ref{fig:Widefield_imaging} shows the image reconstruction results based on recovered TM using FCNN and convolutional U-net networks. It can be seen that all three images can be successfully reconstructed based on recovered TMs using both neural network models. Using the IMMAE (image MAE metric defined in Eqn. \ref{eq:widefield_IMMAE}) we achieved error $\leq 9\%$ and using SSIM (structural similarity index measure metric defined in Eqn. \ref{eq:widefield_SSIM} we achieve similarity $\geq 83\%$. Because of the superior error performance of the FCNN approach, for the remainder of this work we choose this model to demonstrate proof-of-principle performance of neural network TM recovery under a range of realistic scenarios.  However, when scaling to larger size TMs, the convolutional U-net may be more favourable due to fewer trainable parameters.

\subsection{Effect of fiber perturbation widefield image reconstruction}
\begin{figure}[!htpb]
    \centering
    \includegraphics{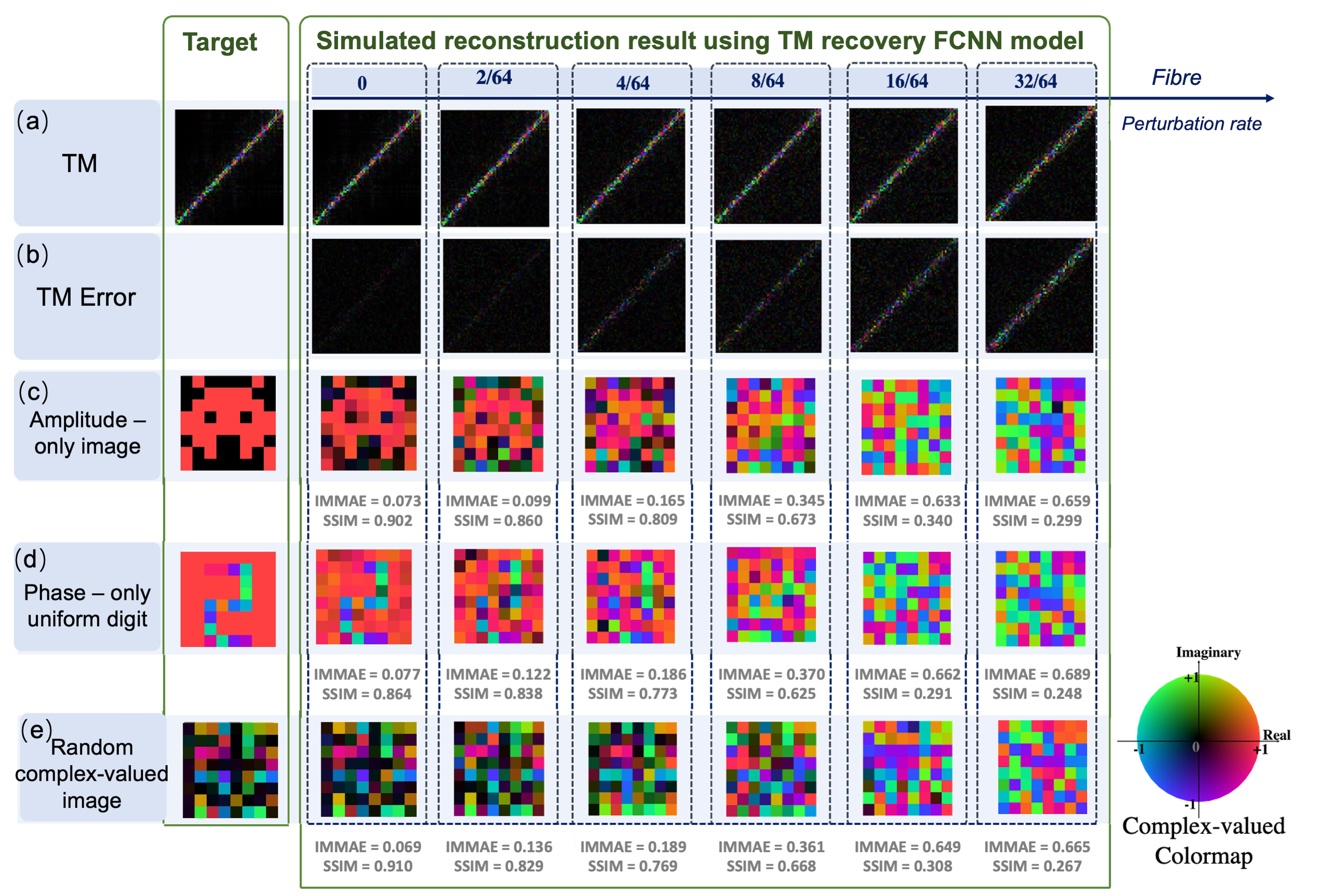}
    \caption{Effect of simulated perturbations of fiber transmission matrices (TMs) during reflection-mode characterization. Perturbations are simulated by swapping columns between two different TMs at wavelength $\lambda_3$. Two metrics are defined to evaluate the image reconstruction performance based on recovered TM using fully-connected neural network (FCNN), namely image mean absolute error, IMMAE (Eqn. \ref{eq:widefield_IMMAE}) and structural similarity index measure, SSIM (Eqn. \ref{eq:widefield_SSIM}). a) Estimated perturbed TM. b) Error in estimated TM. c) Image reconstruction of amplitude-only target (uniform phase) using estimated perturbed TM. d) Image reconstruction of phase-only target (uniform amplitude) using estimated perturbed TM. e) Image reconstruction of complex amplitude and phase target using estimated perturbed TM.In complex-valued colormap, hue represents phase and lightness/darkness represents amplitude.}
    \label{fig:fiber_perturb}
\end{figure}

We then evaluate the robustness of our TM recovery model by simulating the effect of the TM changing mid-way through fiber characterisation. Specifically, we simulate a situation in which the reflection matrices measured at the first two wavelengths ($\mathbf{C_1}$ and $\mathbf{C_2}$) are recorded using a fixed TM, but the reflection matrix measured at the third wavelength $\mathbf{C_3}$ is recorded initially using this same TM but the final few columns are generated using a different TM.  Therefore, we simulated 10 sets of $64\times64$ reflection matrices with five different perturbation rates indicating the numbers of columns swapped (2/64, 4/64, 8/64, 16/64, and 32/64). Figure \ref{fig:fiber_perturb} shows the results of recovering the TM based on these perturbed measurements, and then using this perturbed TM estimate to perform widefield image reconstructions. This is repeated for different fiber perturbation rates based on our pre-trained TM recovery FCNN model. It can be seen that our TM recovery model is compatible with optical fibers with a small perturbation rate (below 6\%) with TM average loss $\leq 8\%$ (standard deviation $\leq 0.82\%$), IMMAE $\leq 19\%$ and SSIM $\geq 76\%$ but performance degrades above this.
\subsection{Confocal image reconstruction based on recovered TM}
\begin{figure}[!h]
    \centering
    \includegraphics{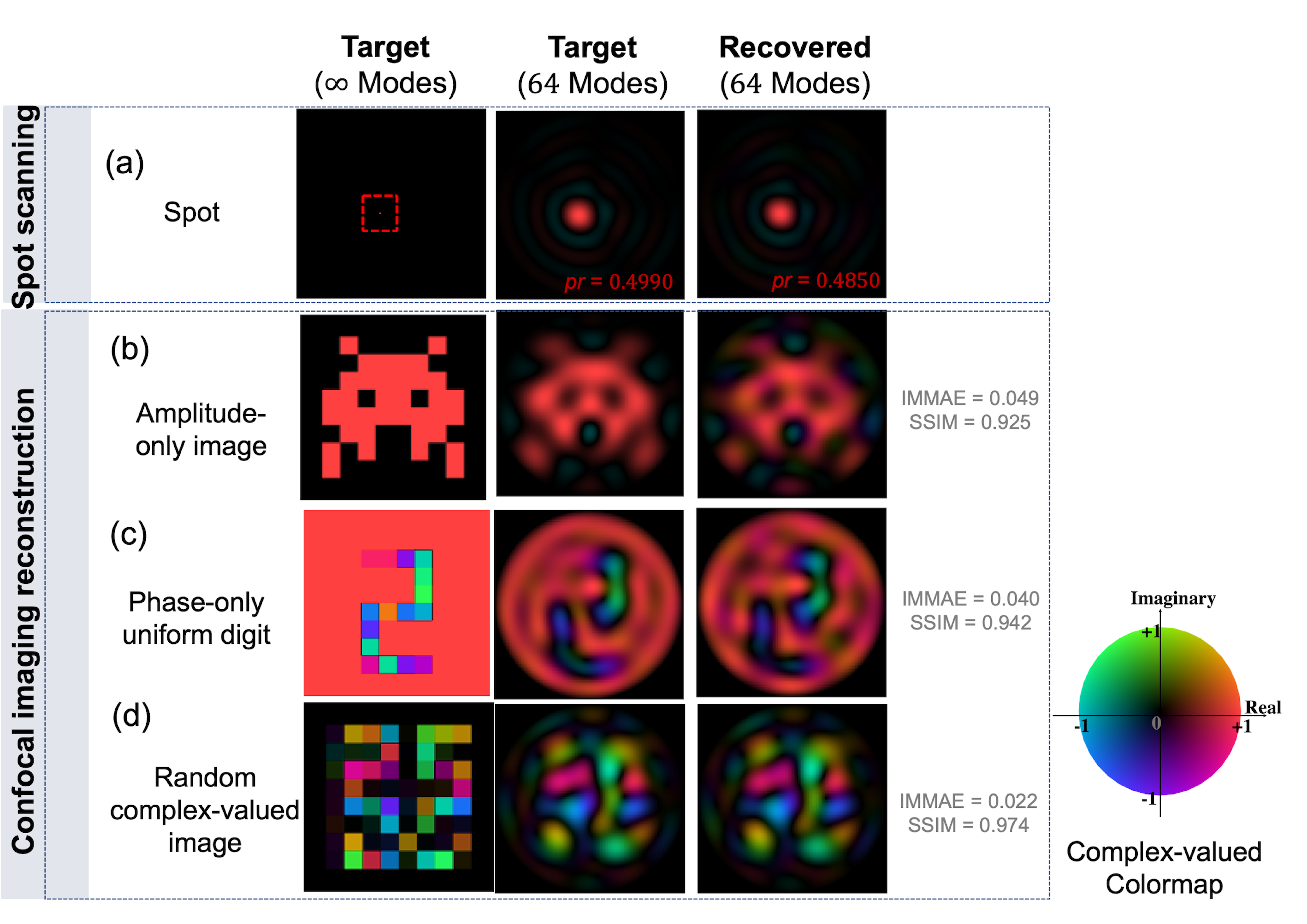}
    \caption{Spot scanning and confocal image reconstruction based on recovered transmission matrices (TMs) using fully-connected neural network (FCNN) model. The spot and 2D full samples are generated using $128\times128$ targets. Power ratio ($pr$) is defined to calculate the average percentage of power in the focus by scanning $128\times 128$ approximate spot positions. Two metrics are defined to evaluate the image reconstruction performance, namely image mean absolute error, IMMAE (Eqn. \ref{eq:widefield_IMMAE}) and structural similarity index measure, SSIM (Eqn. \ref{eq:widefield_SSIM}). a) Creation of a focussed spot in perfect scenario and band-limited scenario with 64 modes using both actual (target) and recovered TMs. b) Image reconstruction using confocal scanning for amplitude-only image, c) Phase-only image and d) Random complex valued image.In complex-valued colormap, hue represents phase and lightness/darkness represents amplitude.}
    \label{fig:Confocal_scanning}
\end{figure}

Many fiber imaging systems use a confocal imaging approach, rather than widefield, because of its superior resolution.  We therefore simulated confocal scanning imaging reconstruction through our recovered TM. For this, we used a simulated LP basis based on a typical fiber profile (see Methods). We addressed $128\times 128$ spot positions and found the average percentage of power in the focus to be 48.5\% using recovered TM, which compares favourably to that achieved using an identity matrix as the TM (49.9\%) as shown in Figure \ref{fig:Confocal_scanning}(a).
Next, we reconstructed confocal images of three target images, denoted $\mathbf{x}\in\mathbb{C}^{128\times 128}$ that contain the same pattern shown in Result Section Widefield image reconstruction based on recovered TM by integrating total reflected power for each spot position. We used $\lambda_4 = 845.5$nm here for imaging. Figure \ref{fig:Confocal_scanning}(b)-(d) shows the confocal image results using FCNN model. It can be seen that all three types of confocal images can be successfully reconstructed based on recovered TMs, with IMMAE $\leq 5\%$ and SSIM $\geq 90\%$. Also, it shows $3\%$ points less image error and $4\%$ points higher similarity between reconstructed and target confocal images compared to widefield imaging.
\subsection{Non-square TM recovery}
\begin{figure}[h]
    \centering
    \includegraphics{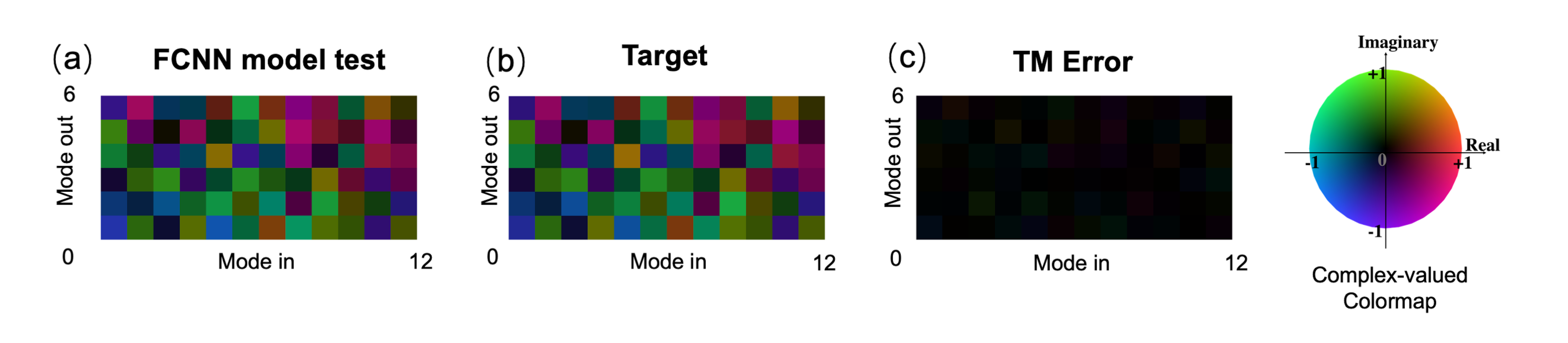}
    \caption{Non-square shaped transmission matrix (TM) $\in \mathbb{C}^{6\times 12}$ recovered by our TM recovery model using fully-connected neural network (FCNN) architecture. a) TM recovered by FCNN model. b) Target TM. c) Error between recovered and target TMs. In complex-valued colormap, hue represents phase and lightness/darkness represents amplitude.}
    \label{fig:Non_squared}
\end{figure}

We next examine the practical case of non-square TMs, e.g. where the desired representation at the distal end of a fiber might be different from that used at the proximal end and may have more elements. This might be the case, for example, when measuring forward TMs to use to train such a network. To recover a TM $\mathbf{A} \in \mathbb{C}^{M_d\times M_p}$, we require that the reflection matrix, $\mathbf{C}\in\mathbb{C}^{M_p\times M_p}$ and that the reflector matrix, $\mathbf{R}\in\mathbb{C}^{M_d\times M_d}$. ${M_p}$ and ${M_d}$ represent the number of elements used for the basis representation at the proximal and distal ends of the fiber respectively. Figure \ref{fig:Non_squared} shows one example of recovered non-square-shaped TM $\in \mathbb{C}^{6\times 12}$ using FCNN, with the average loss of 3.96\% (standard deviation 0.38\%).  `Wide' TMs (with $M_d < M_p$) may be over-constrained by larger dimension $\mathbf{C}$ producing stable solutions, while `tall' TMs (with $M_p < M_d$) may be underconstrained producing degenerate solutions.

\subsection{Reflector matrix conditioning}
Explicit conditions on the reflector matrices, $\mathbf{R}$, namely having distinct eigenvalues, have been a requirement in previous approaches \cite{gordon2019characterizing, Gu2015}. Here, we test our TM recovery FCNN model with different reflector matrix conditioning by varying the number of distinct eigenvalues, as shown in Supplementary Figure 1.  We examined the recovered TMs for 3 conditionings of $\mathbf{R}$: 1 distinct eigenvalue (unitary, all eigenvalues the same), 2 distinct eigenvalues and 6 distinct eigenvalues. The average recovery loss achieved is 5.02\% (standard deviation 0.43\%), 4.96\% (standard deviation 0.41\%), 3.88\% (standard deviation 0.37\%) respectively, demonstrating that the recovery process is compatible with various reflector matrices conditions.

\subsection{Computational resource usage}
\begin{figure}[!h]
    \centering
    \includegraphics{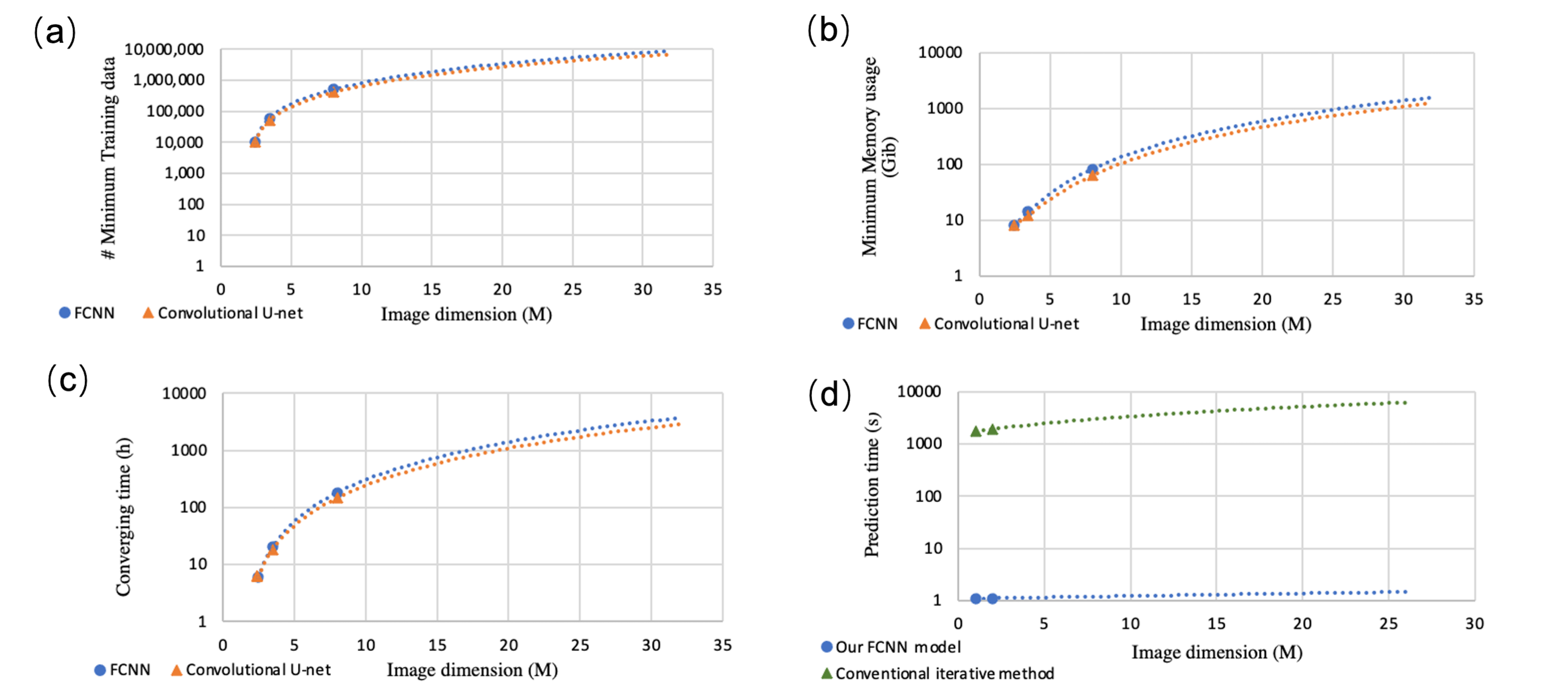}
    \caption{Empirically measured computational resources plotted in log-scale. (a) Minimum training data versus the number of image dimensions, plotting in log-scale.(b) Minimum memory usage versus the number of image dimensions, plotting in log-scale. (c) Converging time versus the number of image dimensions. (d) Prediction time of using our transmission matrix (TM) recovery model and conventional method, plotting in log-scale.}
    \label{fig:Computational}
\end{figure}

 As the dimension of recovered images increases, we expect an increase in the TM dimension thus requiring more computational resources.  Empirically measured computational resources are plotted in log-scale in Figure \ref{fig:Computational} (a)-(c): minimum training data, minimum memory usage, and converging time respectively.  All indicate a quadratic relationship to the image dimension $M$ for both FCNN and convolutional U-net models. For practical imaging applications we would desire at least $32\times32$ image resolution, giving a $1024\times1024$ TM, which would require training with $>$10 million examples, leading to memory consumption $>$1.5TB for the FCNN.  By comparison, the convolutional U-net would require only 1.1TB of memory consumption.  Compared to FCNN, convolutional U-net shows potential advantages in using 25\% fewer memory resources and 20\% less training data within 15\% less training time. Figure 
 \ref{fig:Computational}(d) compares the prediction time using our neural network model with the conventional methods using iterative optimization approaches \cite{gordon2019characterizing}, where our FCNN model shows less reconstruction time ($\sim$1s vs. 1920s for a 12 $\times$ 12 TM). 
%% Experimental TMs test on pre-trained models

\subsection{Validation with experimental TMs}
Finally, to demonstrate applicability to practical systems we cross validated our model on experimentally measured TMs recorded under a representative range of likely fibre conformations \cite{wen2023single}. The graph of loss vs. epoch for the different training steps used to adapt the model pre-trained on simulated data is shown in Supplementary Figure 2. The model is first re-trained from step 1 with additional inputs using random matrices to avoid overfitting to the prior distribution of simulated matrices (Step 2). We next continued training the model on a small subset of the experimental data to aid domain transferability (Step 3). By creating `submatrices' we are effectively implementing a downsampling basis change and so expect some change in correlations betweens TM elements.  However, by including random matrices in our training set we actively prevent the model from learning a strong prior distribution over these correlations and thus overfitting.

The final model achieves recovery of simulated TMs with an average loss of 3.38\% (standard deviation 0.48\%) and experimental matrices with 3.42\% (standard deviation 0.57\%), suggesting that it is applicable to realistic fibre TMs under various conformations. An example of experimental TM used for the test at each step is shown inset in Supplementary Figure 2, where the predicted TM is getting closer to the target TM from Step 1 to Step 3.

\section{DISCUSSION}
We have demonstrated the successful reconstruction of forward fiber TMs based on reflection-mode measurements at multiple wavelengths using a neural network-based approach encompassing two architectures: a fully-connected neural network and a convolutional U-Net.  Previous work applying neural networks to fibers has focussed on image reconstruction as the end goal, but we instead focus on TM reconstruction.  Such an approach is more flexible as the inputs to the network are calibration measurements that reflect a fiber's deformation state at any given time -- previous image reconstruction approaches have instead learned a static representation of the fiber TM encoded in the neural network weights.  Using our approach, the recovered TM will be accurate up to the most recent calibration measurements and can be used for high-speed image recovery via conventional matrix operations.

One major challenge of recovering the TM is the presence of degenerate global phase shift.  Previous work on image reconstruction has addressed this problem by training separate networks for amplitude and phase recovery in purely real space and accepting relatively poor performance for phase recovery \cite{rahmani2018multimode}. Here, we present a loss function that is insensitive to this global phase degeneracy and show a high degree of convergence compared to conventional MAE metrics.  We believe this metric could also find applications in neural network based computer-generated holography or phase retrieval. 

There are several major advantages to our neural network approach compared to previous iterative approaches \cite{gordon2019characterizing}.  First, the prediction time is very fast (typically $\sim$1s), over 4500x faster than the existing iterative approach, which makes this a feasible approach for future real-time imaging,   Training the network is much slower, but this only need to be done once per fiber for a fixed reflector as a one-off calibration step.  Second, our approach shows robustness to the fiber TM changing part way through characterization measurements, as is likely to happen during real \emph{in vivo} usage, and can tolerate up to 6\% column swaps between different reflection matrices.  This performance could be further improved by re-training the network with perturbed examples as input, allowing it to learn an `error correction' strategy. Third, the approach can reconstruct non-square TMs.  This is important because the sampling basis of the light on the proximal facet is often the pixel basis of the camera but at the distal end it may be a mode basis of the fiber (e.g. LP modes). The fibre may support many fewer modes than the camera has pixels. Therefore, to optimize speed and imaging performance it is often desirable to retrieve a TM in the mode basis of the fiber that can easily be addressed using our camera coordinates: hence a non-square TM. Finally, our recovery process is compatible with various reflector matrices conditions whereas in previous models the eigenvalues are required to be distinct \cite{gordon2019characterizing, Gu2015}.  However, it is noted that eigenvalues are unlikely to be identical in any experimental system due to small mode-dependent power losses, but the removal of this condition may increase robustness to noise.

Several trade-offs should be considered when using this approach. The first trade-off is the need for large amounts of experimental transmission and reflection measurements to train the network. This can be addressed by augmenting experimental data with large amount of simulated fiber data: we have previously found good agreement between simulated and experimental matrices \cite{gordon2019characterizing}. To implement these simulations, the reflector matrices would need to be known in advance e.g. by performing TM measurements of a fiber, attaching a reflector, then measuring reflection matrix with minimal fiber perturbation. Alternatively, it may be possible to devise a method of reliably manufacturing reflectors with consistent and highly reproducible properties. Our simulations assume limited fiber bandwidth to avoid degeneracies arising from matrix logarithms, but this bandwidth could be extended using a nonlinear model of the fiber TM over a much broader wavelength range \cite{lee2023efficient}. This joint simulation-experiment approach borrows ideas from data augmentation and domain transfer \cite{wirkert2017physiological, osman2022training, shorten2019survey}. The use of adaptive loss functions, such as in generative-adversarial networks, may further enable convergence using relatively small experimental datasets, or else help to generate further training data.  In these latter scenarios it could be envisioned that the reflector need not be characterized in advanced and can be inferred from a small number of experimental measurements.

The second key trade-off is that the training process is very memory-intensive for dealing with large sizes of TM that are typically encountered in imaging applications e.g. $1024\times1024$ would require over 1TB for training the recovery model. The convolutional U-net architecture has much ($\times$1000) fewer trainable parameters, which reduces memory usage somewhat, although the large dimension of each input TM also has a large influence on the memory usage.  However, using convolutional architectures may come at the expense of slightly increased error (a few percent) as we observed in our TM reconstruction results. One possible solution is to develop matrix compression techniques such as auto-encoder models to represent matrices in a low-dimensional latent space. Reducing batch size will also reduce memory usage but can lead to greater fluctuations and poorer convergence.

We anticipate this neural network-based TM recovery model will lead to machine-learning models for complex-valued data, for example in holographic imaging and projection and phase retrieval, where both phase control and speed are required.

\section{CONCLUSIONS}
Overall, our model for reconstructing $64\times64$ complex-valued fiber TMs through a single-ended optical fiber system achieves $\leq 4\%$ error for both FCNN and convolutional U-net based neural network architectures. By re-training our FCNN model we also show that it can reconstruct experimentally measured $64\times64$ TMs with $3.42\%$ error. Using our recovered TMs we demonstrate image reconstruction of complex objects in two imaging modalities: 8$\times8$ pixel widefield imaging, achieving IMMAE $\leq9\%$ and SSIM $\geq83\%$, and confocal scanning, achieving $48.5\%$ power focused in the spot and $\leq5\%$ IMMAE and $\geq90\%$ SSIM.

\section{METHODS}
\subsection{Simulated data generation}
\begin{figure}[!htbp]
    \centering
    \includegraphics{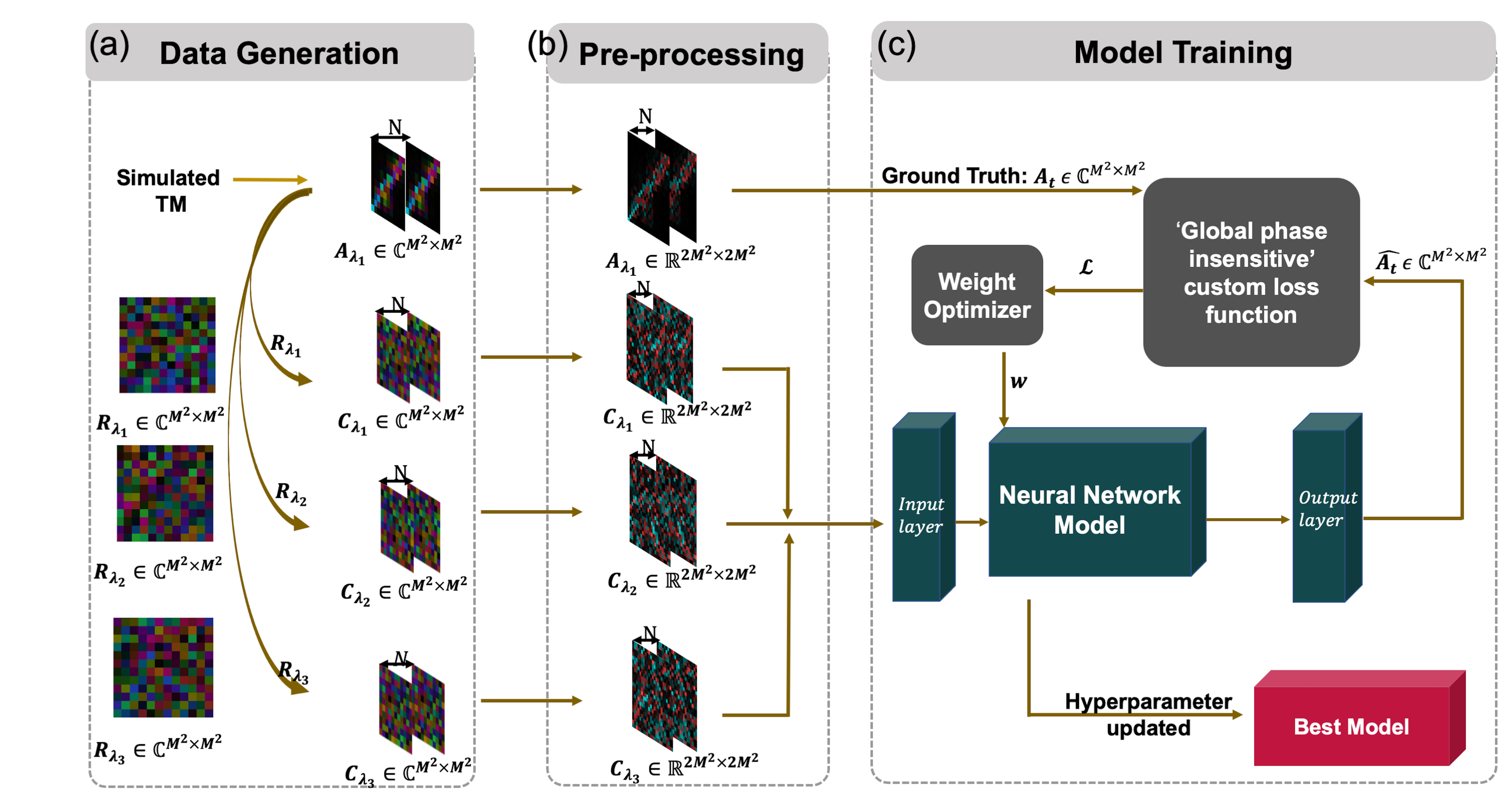}
    \caption{Schematic of transmission matrix (TM) recovery model. This contains (a) data generation, (b) data pre-processing and (c) model training. $N$ pairs of TM are firstly simulated as the ground truth. $\lambda_1$, $\lambda_2$ and $\lambda_3$ represent three different wavelengths (in our case, 850nm, 852nm and 854nm). The input of the model is all real-valued matrices concatenated with reflection matrices at three different wavelengths. $\textit{L}$ represents custom loss function and $\textit{w}$ represents the weight updated by the optimizer.}
    \label{fig:NNprocess}
\end{figure}

We present a TM recovery method that uses neural networks, instead of using iterative approaches \cite{gordon2019characterizing}, to solve the Eqn. \ref{eqn2} -- Eqn. \ref{eqn4}. Figure \ref{fig:NNprocess} shows the schematic of this TM recovery model. Specifically, we first simulated $N$ optical fiber TMs, $\mathbf{A_{\lambda_1}}\in\mathbb{C}^{M^2\times M^2}$, at a wavelength of ${\lambda_1}$ as the ground truth. Here we use $M=8$ and $\lambda_1=850nm$. Our simulation model aims to recreate typical properties found in practical fiber TMs.  First, we assume that TMs are sparse in some commonly used basis e.g. LP modes for multimode fibers or pixel basis for multicore fibers \cite{gordon2019coherent}. Second, we assume TMs can be arranged such that the majority of power intensity lies along the main diagonal with additional power spread along sub-diagonals, which is also typically observed when using bases closely matched to the fiber eigenbasis\cite{carpenter2014110x110}. Third, we assume that realistic TMs are slightly non-unitary, with mode-dependent loss values (i.e. condition numbers) in the range of 3-5 \cite{carpenter2014110x110, gordon2019coherent}. To meet these assumptions, we generate a uniformly distributed random tri-diagonal matrix, $\mathbf{B}\in\mathbb{C}^{64\times 64}$, which has non-zero elements only at the main diagonal, subdiagonal and superdiagonal. We then compute the left singular matrix $\mathbf{U}\in\mathbb{C}^{64\times 64}$ and right singular matrix $\mathbf{V}\in\mathbb{C}^{64\times 64}$ via singular value decomposition (SVD). We then apply a new singular value distribution via a matrix $\mathbf{S_{new}}\in\mathbb{R}^{64\times 64}$, a diagonal matrix that contains random values along its diagonal ranging from 0.5 to 2.5 to simulate our expected TM. We then construct the TM at $\lambda_1$ as:
\begin{equation}
    \mathbf{A_{\lambda_1}} = \mathbf{U}*\mathbf{S_{new}}*\mathbf{V}^T
\end{equation}
Next, we apply Eqn. \ref{eqn2} - Eqn. \ref{eqn4} to simulate corresponding TMs, $\mathbf{A_{\lambda_1}}$ and $\mathbf{A_{\lambda_1}}$ at wavelengths $\lambda_2=852nm$ and $\lambda_3=854nm$. Here, we use wavelengths $\lambda_1=850$nm, $\lambda_2=852$nm and $\lambda_3=854$nm as physically realistic values within the TM bandwidth of a typical endoscope length fiber ($\sim$2m) \cite{gordon2019characterizing}.

Following this, we simulate reflectors on the fibre distal tip by generating three complex-valued matrices with complex uniform randomly distributed elements (-1 to 1 and -$i$ to $i$). These three matrices then form our reflector matrices, $\mathbf{R_{\lambda_1}}\in\mathbb{C}^{M^2\times M^2}$, $\mathbf{R_{\lambda_2}}\in\mathbb{C}^{M^2\times M^2}$, and $\mathbf{R_{\lambda_3}}\in\mathbb{C}^{M^2\times M^2}$ at wavelengths ${\lambda_1}$, $\lambda_2$ and $\lambda_3$ respectively. By generating reflectors in this way we ensure with high probability, according to random matrix theory, that the eigenvalues are distinct \cite{arnold1971wigner}.

Finally, we combine the TMs and reflector matrices to generate $N$ sets of complex-valued reflection matrices $\mathbf{C_{\lambda_1}}\in\mathbb{C}^{64\times 64}$, $\mathbf{C_{\lambda_2}}\in\mathbb{C}^{64\times 64}$, and $\mathbf{C_{\lambda_3}}\in\mathbb{C}^{64\times 64}$ at three different wavelengths $\lambda_1=850$nm, $\lambda_2=852$nm and $\lambda_3=854$nm.

To feed this data to our neural network, which only accepts real numbers, we convert inputs and outputs from complex to real-valued data. A $2\times2$ complex-valued matrix can be represented by a $4 \times 4$ real-valued matrix as follows \cite{hile1990matrix}:
\begin{equation}
\label{eq18}
    \begin{bmatrix}
    {a+bi}&{c+di}\\
    {e+fi}&{g+hi}
    \end{bmatrix} \mapsto
    \begin{bmatrix} 
    a&{-b}&c&{-d}\\
    b&a&d&c\\
    e&{-f}&g&{-h}\\
    f&e&h&g\\
    \end{bmatrix} 
\end{equation}
\noindent where, $\mapsto$ indicates an isomorphism.

Finally, the input of the model, three $\mathbf{C_{\lambda}}\in\mathbb{C}^{128\times 128}$ at different wavelengths are normalized using in the range from -1 and 1. Each set of 3 reflection matrices, $\mathbf{C_{\lambda_{1..3}}}$ represents a single input to our neural network model. We split the $N$ data into training, validation and test sets with an 8:2:1 ratio. The validation set is expected to provide unbiased evaluations and stopping criteria on unseen data, and test set aims to examine the generalization performance of the model on unseen data. Test set is independent to the validation set and contains unseen data that is not used during training process. The model is trained with the ADAM optimizer using our custom-defined loss function. Python was used for model training and MATLAB was used for data pre-processing and post-processing because of its ease of use for complex matrix computations.

\subsection{Imaging modalities}
We validate our system via two commonly used fiber imaging modalities: wide-field and confocal scanning.  For wide-field imaging we assume a pixel basis at the fiber distal end, and that a ground-truth complex image $\mathbf{X}\in \mathbb{C}^{M^2}$ passes via the simulated optical fiber ground-truth TM at a fourth wavelength $\lambda_4=854.5$nm. We then use the recovered TM at a fourth wavelength $\lambda_4=854.5$nm to estimate original image and compare this to the ground-truth. For simplicity, we assume the object is located at the distal fiber facet with no loss coupling into and out of the fibre, and we neglect the loss in transferring through the reflector stack but consider random Gaussian noise to better simulate expected behaviour in a real system. Theoretically, the reconstructed widefield image, $\mathbf{\hat{X}_j}\in\mathbb{C}^{M^2}$ can be calculated by:
\begin{equation}
    \mathbf{\hat{X_j}} =\mathbf{(\hat{A}_{\lambda_4}}^T)^{-1}(\mathbf{{A_{\lambda_4}}}^T\mathbf{X_j}+\mathbf{Z})
    \label{eq:imaging}
\end{equation}
\noindent where, $\mathbf{Z}$ is random Gaussian noise with a power of 2\% of the target signal. $\mathbf{\hat{X}_j}$ is the reconstructed image, $\mathbf{X_j}$ is the target image, $\mathbf{\hat{A}_{\lambda_4}}$ and $\mathbf{A_{\lambda_4}}$ are the recovered TM and target TM at wavelength $\lambda_4$ respectively.

For confocal imaging, which is commonly used in fiber imaging \cite{Loterie2015}, we assume an LP mode basis at the fiber output, which is a realistic and widely used fiber basis \cite{ploschner2015seeing}. To implement this, we first simulated a multimode fiber with a core radius of $30\mu m$, length of $1.5m$ and a numerical aperture (NA) of 0.24 and used 64 of the available modes as our basis for creating a spot to be scanned. We used our recovered TM to estimate the required proximal field to create a spot then examined the expected spot by forward propagating via the ground truth TM. We next scanned $128\times128$ approximate spot positions and found the average percentage of power within the FWHM of the spot. Using this approach we reconstructed confocal images of three target images, denoted $\mathbf{x}\in\mathbb{C}^{128\times 128}$ that contain the same pattern shown in Result Section Widefield image reconstruction based on recovered TM by integrating total reflected power for each spot position. 

\subsection{Metrics}
We next define loss metrics to compare the complex overlap integral between the target and recovered results. However, due to the degenerate global phase factor, this complex overlap integral is multi-valued.  Therefore, the loss metrics defined here first normalize to correct for the global phase term (see Methods Section Global phase insentive loss function) before calculating the complex correlation. This is an essential step to gauge the accuracy both for TM recovery and image reconstruction in widefield and confocal modalities. 

To gauge the accuracy of our TM recovery we define an average loss metric by calculating the average mean absolute error (MAE) of each validated TM in the test data:
\begin{equation}
    \mathrm{average\ loss\ (\mathbf{\hat{A}_t},\mathbf{A_t})} = \frac{1}{N_{test}}\times\frac{1}{4M^2}\sum_{n=1}^{N_{test}}\sum_{t=1}^{4M^2}|\mathbf{\hat{A}_t}-\mathbf{A_t}e^{i\phi_c}|
    \label{eq:averageloss}
\end{equation}
\noindent where, $\mathbf{\hat{A}_t}$ is the recovered TM, $\mathbf{A_t}$ is the target TM, $e^{i\phi_c}$ is the global phase normalization term, $4M^2$ is the total number of TM elements, and $N_{test}$ is the number of data used for testing.

To gauge the accuracy of widefield reconstructed image based on the recovered TM, we define two metrics, namely the image MAE (IMMAE) and complex-valued based structural similarity index measure (SSIM) of each reconstructed image:
\begin{equation}
    \mathrm{IMMAE\ (\mathbf{\hat{X}_j}, \mathbf{X_j})} = \frac{1}{M^2}\sum_{j=1}^{M^2}|\mathbf{\hat{X}_j}-\mathbf{X_j}e^{i\phi_c}|
    \label{eq:widefield_IMMAE}
\end{equation}
\begin{equation}
    \mathrm{SSIM\ (\mathbf{\hat{X}_j}, \mathbf{X_j})} = \frac{2|\sum_{j=1}^{M^2}\mathbf{\hat{X}_j}\mathbf{X_j}^*e^{i\phi_c}|+K}{\sum_{j=1}^{M^2}|\mathbf{\hat{X}_j|}^2+\sum_{j=1}^{M^2}|\mathbf{X_j|}^2+K}
    \label{eq:widefield_SSIM}
\end{equation}
\noindent where, $M^2$ is the total number of pixels of image $\mathbf{X_j}$, $\mathbf{X_j}^*$ is the complex conjugate of $\mathbf{X_j}$, $K = 0.03$ is a positive constant to improve the robustness when the local signal to noise ratios are low. These two metrics can also be extended to evaluate the performance of recovered confocal images. 

In the confocal case, we compare recovered images $\mathbf{\hat{Y}}$ against the confocal imaging scenario with an identity TM, $\mathbf{Y}$, using the two metrics (IMMAE and SSIM) introduced in Eqn. \ref{eq:widefield_IMMAE} and Eqn. \ref{eq:widefield_SSIM}. This aims to do a fair comparison considering the effect of conversion between fiber mode basis and pixel basis.

\subsection{Network architectures}
\begin{figure}[!htbp]
    \centering
    \includegraphics{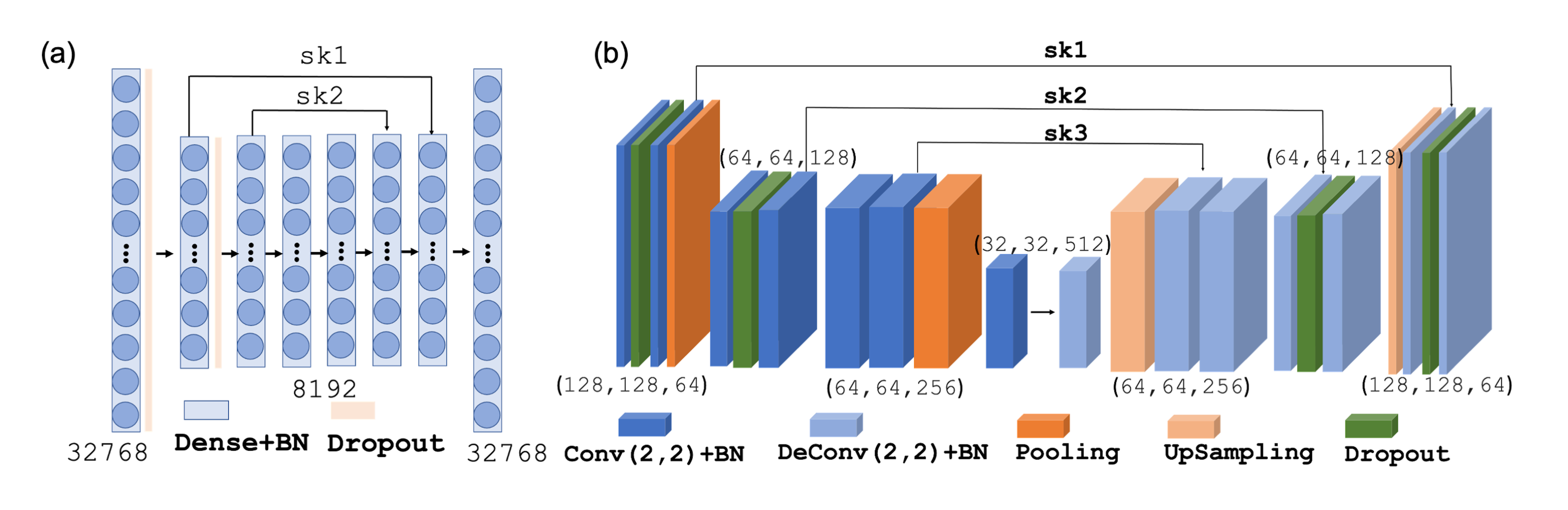}
    \caption{Architectures of two different neural network models used for TM recovery. (a) Fully-connected neural network, (b) Convolutional U-net}
    \label{fig:NNarch}
\end{figure}

We defined two neural network models: a fully-connected neural network (FCNN) and convolution U-net-based neural network as shown in Figure \ref{fig:NNarch}. The FCNN is a ten-layer densely connected neural network (eight hidden layers), including 32,768 neurons in first and last hidden layers and 8192 neurons in other layers, all with Leaky ReLU activation function. Figure \ref{fig:NNarch} shows the FCNN architecture, where reflection matrices $\mathbf{C_{\lambda_1}}\in\mathbb{R}^{128\times 128}$, $\mathbf{C_{\lambda_2}}\in\mathbb{R}^{128\times 128}$, $\mathbf{C_{\lambda_3}}\in\mathbb{R}^{128\times 128}$ are flattened into $1D$ arrays and then concatenated as the input of the model (with the size of $49152\times1$) and TM $\mathbf{A_{\lambda_1}}\in\mathbb{R}^{128\times 128}$, flattened into $1D$ array as the output (with the size of $16384\times1$). Batch normalization layers were defined between every dense layer and dropout layers at the rate of $0.2$ were defined after the first two dense layers. Also two skip connections were developed in order to prevent the model overfitting. The model was trained iteratively with the global phase-insensitive custom loss function used. The training dataset for recovering $64\times64$ TM consisted of 500,000 matrices and the model was run for 2500 epochs, taking 182.5 hours using Tensorflow 2.0 running on a NVIDIA Tesla V100 GPU. The Adam optimizer was used with a learning rate of 0.004 in a decay rate of $1e^{-4}$.

Next, we developed a U-net-based model that used encoder-decoder architecture, including seven Conv2D and DeConv2D layers respectively and two MaxPooling and UpSampling layers respectively with LeakyRelu activation function in each layer. Figure \ref{fig:NNarch} shows this architecture, where reflection matrices $\mathbf{C_{\lambda_1}}\in\mathbb{R}^{128\times 128}$, $\mathbf{C_{\lambda_2}}\in\mathbb{R}^{128\times 128}$, $\mathbf{C_{\lambda_3}}\in\mathbb{R}^{128\times 128}$ are defined in three channels as the input of the model (with the size of $128\times128\times3$) and TM $\mathbf{A_{\lambda_1}}\in\mathbb{R}^{128\times 128}$, as the output (with the size of $128\times128\times1$). Batch normalization layers were defined between every layer and dropout layers at the rate of $0.2$ were defined after the second and last second Conv layers. Also three skip connections were developed in order to prevent the model being overfitted. The model was trained iteratively with the global phase insensitive custom loss function defined. Also, 2200 epochs were used for training 400,000 training datasets using 143h. The Adam optimizer was used with a learning rate of 0.004 in a decay rate of 1e-4.

\subsection{Global phase insensitive loss function}
\label{subsec:phaseinsensitiveloss}
Widely-used conventional loss functions such as mean absolute error (MAE) or mean squared error (MSE) calculate the absolute difference between predicted and target output values. However, there is a class of problems whose solutions trained by deep learning models are degenerate within a global phase factor, but whose relative phase between pixels must be preserved. This class includes problems where complex TMs are reconstructed and relative phase, but not global phase, could extend to phase-hologram generation algorithms where replay-field phase is relavant \cite{zheng2023global}. This is depicted visually in Supplementary Figure 3(a), which shows one example of a pair of predicted and target matrices with complex entries depicted as vectors. Supplementary Figure 3(b) shows the complex error between these two matrices when using MAE as the loss function. Due to the global phase shift, we observe that the vectors have large magnitudes, which will lead to an overall very large MAE when their magnitudes are summed. In the limiting case (e.g. when the phase shift is of $\pi$) where the predicted and target matrices are identical, this global phase shift can result in a normalized MAE of 100\% when the true value should be 0\%. To avoid this problem, we propose a custom loss function termed a `global phase insensitive' loss function that normalizes for this global phase factor:
\begin{equation}
    L(\mathbf{\widehat{A_t}}(w),\mathbf{A_t}) = \sum_{t=1}^{4M^2}\left |  \mathbf{\widehat{A_t}}(w)-\mathbf{A_t} e^{i \phi (\mathrm{Tr}(\mathbf{A_t}^H\mathbf{\widehat{A_t}}(w))}\right | + \frac{\alpha}{2} {\|w\|}^2
    \label{eqn:loss3}
\end{equation}
\noindent where, $\mathbf{\widehat{A_t}}(w)\in\mathbb{C}^{M^2\times M^2}$ and $\mathbf{A_t}\in\mathbb{C}^{M^2\times M^2}$ represent predicted and target output value with regards to weight, $w$, respectively, $\sum$ represents summation over all matrix elements, $\phi$ represents the argument function for a complex number input. We add an $\ell_2$ regularization term to encourage generalization of the model with regularization parameter, $\alpha$ = $10^{-4}$. This formula implicitly weights the phase contributions by the product of magnitudes of the respective elements in $\mathbf{\widehat{A_t}}(w)$ and $\mathbf{{A_t}}$, which upon convergence will approximately equal the squared magnitude of the target.

The rationale for this is that when the optimization algorithm approaches a minimum, the remaining error for each complex element will be entirely due to aleatoric uncertainty, e.g. a circularly symmetric zero-mean complex Gaussian distribution \cite{bhatt2021variational}. To estimate the correction factor, the element-wise complex errors can be summed, as shown in Supplementary Figure 3(c). This will produce an overall complex factor that has the desired global phase shift, shown in Supplementary Figure 3(d). The predicted output value can be corrected by multiplying by this phase factor as shown in Supplementary Figure 3(e), the result of which is then used to compute further parameter updates in the gradient descent algorithm. It can be seen that the complex error in Supplementary Figure 3(f) between the predicted and target output value is reduced to a minimum after removing the phase factor compared to that calculated by MAE. We then compared the absolute values of the complex error calculated by MAE (green bar) and our customized global phase insensitive loss function (blue bar) respectively over 100,000 pairs of predicted and desired TM as shown in Supplementary Figure 3(g). The error using the custom loss function is more than two times smaller than that of the conventional loss function (MAE).

\subsection{Experimental measurement and updated model setup}
The optical layout of our experimental setup is provided in \cite{wen2023single}. Measuring the TM, treated as a linear operator, involves input modulation and output recovery. To achieve input modulation, the incident light was collimated into MMF by an objective lens and a 4$f$ configuration after reflecting from digital micro-mirror device (DMD). In our specific case, the input fields are implemented using the Hadamard basis, providing an orthogonal set of modulated inputs. For the complex output recovery, images of the output facet of the fiber are captured using a movable calibration module. The light from the multimode fiber (MMF) is combined with the reference signal using a beam splitter. The reference light is directed to the camera through a single-mode fiber, and then the off-axis holography is recorded by CCD camera after magnifying by a microscope objective. By employing this optical layout, we are able to accurately measure the TM and gain insights into the propagation characteristics of light through the MMF.  Using this approach, we measured experimental TMs for 164 different bending conformations and selected 64 rows and 64 columns of the raw measurements to create suitably sized matrices for training our model. TMs are measured at a wavelength of 488 nm and we use these as our base $A_{\lambda_1}$ at wavelength $\lambda_1$. We then scale this matrix to $\lambda_{2-3}$ using the method in Eqn. \ref{eqn3} and Eqn. \ref{eqn4}, which has been validated previously on multi-wavelength TM datasets \cite{gordon2019characterizing}.

For a small size matrix ($8 \times 8$) we find that the model trained on simulation data is sufficient to reconstruct cropped $8 \times 8$ experimental matrices.  However, for the larger $64 \times 64$ matrices, the model pre-trained on simulated data is prone to overfitting as it tends to also learn the prior distribution of TMs, and the subsequent loss is around $9.24\%$ (standard deviation 0.78\%). This learning of a prior can be advantageous in some cases where prior information is essential to resolve degeneracies, but to avoid that here we next re-trained the model with an augmented dataset containing 600,000 additional random matrices with elements drawn from complex uniform distribution.  This reduced loss to an average value of around $5.07\%$ (standard deviation of 0.67\%), but a degree of overfitting is still present and so tweaked the model by continuing the training on 100 experimental TM, keeping 64 aside for validation and testing. Visually, these experimental TMs are structured differently to our simulated TM due to their use of a different basis. However, the training on random matrices seems to encourage generalisation to different matrix structures.

\section*{DATA AVAILABILITY}
The data presented in this study are available from the following source:
https://doi.org/10.17639/nott.7334

\section*{CODE AVAILABILITY}
The code for this study is available from the following source:
https://doi.org/10.17639/nott.7334.

\section*{AUTHOR CONTRIBUTIONS}
YZ: Idea conception, code implementation, result analysis, manuscript writing. TW: Code implementation for confocal imaging. WZ, QY: Collection of experimental data, manuscript writing. GG: Idea conception, project supervision, manuscript writing.

\section*{COMPETING INTERESTS}
The authors declare no competing interests.

\section*{ACKNOWLEDGEMENT}
The authors acknowledge support from a UKRI Future Leaders Fellowship (MR/T041951/1) and an EPSRC Ph.D. studentship.

%Bibliography
\bibliographystyle{unsrt}  
\bibliography{references}  
\end{document}